\documentclass[%
    aps,
    prb,
    reprint,
    superscriptaddress,
    amsfonts,
    amsmath,
    amssymb,
    eprint,
    longbibliography,
]{revtex4-2}

\usepackage[utf8]{inputenc}
\usepackage[T1]{fontenc}

\DeclareUnicodeCharacter{2009}{\ }

\usepackage{silence}
\usepackage{graphicx}
\usepackage{dcolumn}
\usepackage{bm}
\usepackage{mathtools}
\usepackage{physics}
\usepackage{nicematrix}
\usepackage{stmaryrd}
\usepackage{hyperref}
\usepackage{cleveref}
\Crefname{figure}{Fig.}{Figs.}
\Crefname{equation}{Eq.}{Eqs.}
\usepackage{import}
\usepackage{color}
\usepackage{lipsum}
\usepackage[disable]{todonotes}
\usepackage{verbatim}
\usepackage{tikz}
\usepackage[normalem]{ulem}
\usetikzlibrary{arrows.meta}
\usetikzlibrary{backgrounds}
\usetikzlibrary{arrows,decorations.pathmorphing}

\usepackage[caption=false]{subfig}

\usepackage{textgreek}
\usepackage{etoolbox}
\apptocmd{\sloppy}{\hbadness 10000\relax}{}{}

\begin{document}

\title{Subgap Linear Thermoelectricity in Superconducting Quantum Hall Systems}

\author{Leonardo Pierattelli}
\affiliation{Istituto Nanoscienze – CNR, NEST-SNS, Piazza San Silvestro 12, I-56127 Pisa, Italy}
\author{Fabio Taddei}
\affiliation{Istituto Nanoscienze – CNR, NEST-SNS, Piazza San Silvestro 12, I-56127 Pisa, Italy}
\author{Alessandro Romito}
\affiliation{Department of Physics, Lancaster University, Lancaster LA1 4YB, United Kingdom}
\author{Alessandro Braggio}
\affiliation{Istituto Nanoscienze – CNR, NEST-SNS, Piazza San Silvestro 12, I-56127 Pisa, Italy}
\affiliation{Institute for Quantum Studies, Chapman University, Orange, CA 92866, USA
}

\date{\today}

\begin{abstract}
We show that an integer quantum Hall setup proximized by superconductors can exhibit subgap thermoelectric effects in the linear-response regime when triplet superconducting correlations are present.
We devise a minimal setup that enables a nonzero Seebeck effect mediated by Andreev processes and predict that the corresponding Seebeck coefficient can reach values on the order of $k_B/e$ in the middle of the quantum Hall plateau.
We analytically show that both triplet correlations and spin polarization are essential for the emergence of the thermoelectric effect, which arises despite the linear band dispersion of the edge states.
We characterize the dependence of the thermoelectric response on the Hamiltonian parameters and the system's temperature regime.
\end{abstract}

\maketitle

\paragraph*{Introduction.}
Thermoelectricity is the ability of an electronic system to generate a voltage bias in response to a thermal gradient \cite{benenti_FundamentalAspectsSteadystate_2017}.
It is the subject of intense research due to its potential applications in energy harvesting and, more recently, in the enhanced control of heat and charge transport in quantum systems at ultralow temperatures~\cite{zhang_MicrothermoelectricDevices_2022,blasi_NonlocalThermoelectricEngines_2021,battisti_BipolarThermoelectricSuperconducting_2024}. 
In quantum devices, thermoelectric (TE) effects can also be exploited to probe systems that are not easily accessible by conventional techniques~\cite{cai_SensitiveRoomtemperatureTerahertz_2014,lundeberg_ThermoelectricDetectionImaging_2017,wang_ThermalGenerationManipulation_2020,mateos_NonlocalThermoelectricityQuantum_2024,yang_NonlinearThermoelectricEffects_2025}. 
A particularly interesting platform in this context is provided by superconducting (SC) nanodevices, which lie at the heart of a wide range of sensing and computational technologies.
In SC nanodevices, thermoelectricity can arise in the non-linear regime~\cite{sanchez_NonlinearPhenomenaQuantum_2016}, but non-trivial TE effects have also been uncovered in the linear-response regime~\cite{arrachea_ThermoelectricProcessesQuantum_2025,dutta_ThermalTransportSuperconductor_2025}.
These effects generally originate from the breaking of energy-inversion symmetry (EIS) of the scattering coefficients with respect to the superconducting chemical potential \cite{arrachea_ThermoelectricProcessesQuantum_2025}.
Such symmetry breaking can be achieved through several mechanisms: spontaneous particle–hole symmetry breaking~\cite{marchegiani_NonlinearThermoelectricityElectronHole_2020,marchegiani_SuperconductingNonlinearThermoelectric_2020,germanese_BipolarThermoelectricJosephson_2022,bernazzani_BipolarThermoelectricityBilayerGraphene_2023}; spin-splitting fields in hybrid SC systems in the supra-gap regime~\cite{bergeret_ColloquiumNonequilibriumEffects_2018,machon_GiantThermoelectricEffects_2014,machon_NonlocalThermoelectricEffects_2013,ozaeta_PredictedVeryLarge_2014,kolenda_NonlinearThermoelectricEffects_2016,kolenda_ObservationThermoelectricCurrents_2016,gonzalez-ruano_ObservationMagneticState_2023,araujo_SuperconductingSpintronicHeat_2024}; topological and non-local configurations~\cite{blasi_NonlocalThermoelectricitySuperconductorTopologicalInsulatorSuperconductor_2020}; non-trivial SC phases~\cite{trocha_ThermoelectricPropertiesQuantum_2025,guarcello_ThermoelectricSignaturesOrderparameter_2023,hwang_OddfrequencySuperconductivityRevealed_2018,singh_GiantThermoelectricResponse_2024}; and odd-frequency superconductivity in the subgap regime~\cite{dutta_ThermoelectricityCarriedProximityinduced_2020,savander_ThermoelectricDetectionAndreev_2020}

TE properties of quantum Hall (QH) systems have been extensively investigated in  Refs.~\cite{jonson_ThermoelectricEffectWeakly_1984,fletcher_SearchTrendsThermopower_1988,yang_ThermopowerPossibleProbe_2009,barlas_ThermopowerQuantumHall_2012,kobayakawa_DiffusionThermopowerQuantum_2013,real_ThermoelectricityQuantumHall_2020,sheng_ThermoelectricResponseEntropy_2020,braggio_NonlocalThermoelectricDetection_2024}, while the study of TE properties at the interface between QH systems and SCs is more recent~\cite{giazotto_LandauCoolingMetal_2007,panu_HeatchargeSeparationHybrid_2024,zhao_ThermalPropertiesSuperconductorQuantum_2025,mccourt_ThermoelectricEffectQuantum_2025,wang_ThermoelectricDetectionCrossed_2026}. 
Significant progress in fabrication techniques over the past few years has enabled the reliable realization of such hybrid structures, particularly in graphene-based heterostructures~\cite{zhao_InterferenceChiralAndreev_2020,vignaud_EvidenceChiralSupercurrent_2023,indolese_CompactSQUIDRealized_2020,zhao_LossDecoherenceQuantum_2023,zhao_NonlocalTransportMeasurements_2024,wang_AndreevReflectionsNbN_2021,gul_AndreevReflectionFractional_2022,barrier_OnedimensionalProximitySuperconductivity_2024,hatefipour_AndreevReflectionQuantum_2024,hatefipour_InducedSuperconductingPairing_2022,jang_EdgeDependenceSupercurrent_2025}, with potential applications in quantum technologies \cite{lee_GraphenebasedJosephsonJunction_2020}.
However, a full theoretical treatment of the thermoelectric properties of QH-SC setups in the middle of the QH plateau is missing.

\begin{figure}[t]
    \includegraphics[width=\linewidth]{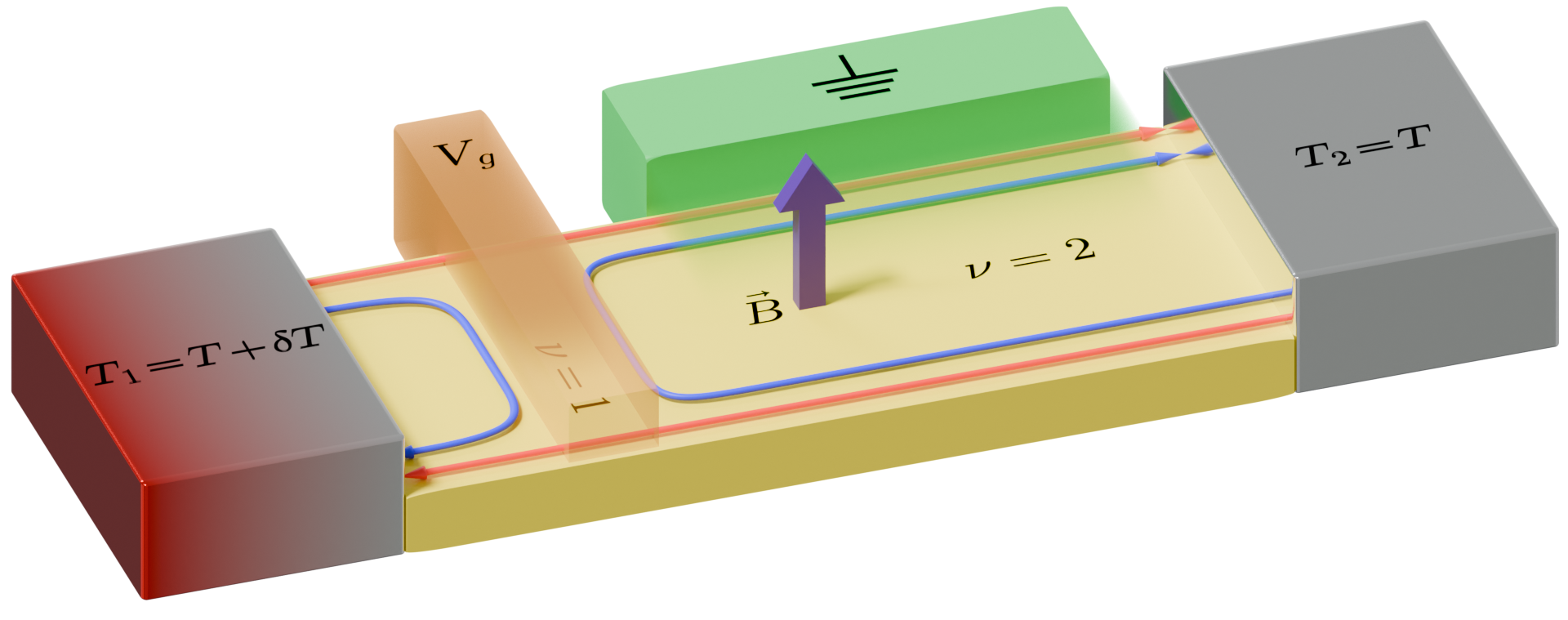}
    \caption{Sketch of the QH bar (yellow) at $\nu=2$, attached to two normal electrically floating and thermally biased metallic contacts.
    Spin-resolved edge states are represented by red and blue lines for spin-up and spin-down states, respectively.
    The top orange gate at voltage $V_g$ depletes the 
    2DEG such that the edge states with spin antiparallel (in blue) to the magnetic field (in purple) are fully reflected.
    The edge modes traveling in the proximized region in front of the grounded superconductor (green) are
    then collected in contact 2
    }
    \label{fig:dev}
\end{figure}

We propose an effective model to describe thermoelectricity in a hybrid QH-SC device at the $\nu=2$ plateau, which, for sufficiently strong Zeeman splitting, supports spin-polarized edge states.
Our main result is that finite subgap linear TE coefficients arise when triplet components are present in the superconducting correlations. This effect emerges uniquely from the interplay between quantum Hall edge states and superconductivity, since the edge states alone do not exhibit thermoelectricity due to their linear band dispersion.
We determine the thermopower $S$ as a function of the length of the superconducting region, of the magnitude and direction of the triplet component, and of the superconducting order-parameter phases. In the special case of a triplet pairing vector aligned with the spin-quantization axis and real superconducting order parameters, we derive an analytical expression for $S$, explicitly revealing the role of the triplet component in generating a finite thermoelectric response.

\paragraph*{Model.}

We analyze the TE properties of a
QH bar shown in \Cref{fig:dev}. Two normal contacts located at the opposite sides of the bar are maintained at different temperatures while remaining electrically floating: the left‑hand (right-hand) side contact is kept at $T_1 = T + \delta T$ ($T_2 = T$).
The Hall bar, redwith a magnetic field $\vec{B}$, is tuned in the middle of the QH plateau where electronic transport is fully dominated by edge states, at filling factor $\nu = 2$, with spin‑resolved edge states.
The green element in \Cref{fig:dev} represents a superconductor, of length $L$, that proximizes the nearby electron gas' edge.
The superconductor is assumed to be grounded, with electrochemical potential $\mu_s = 0$ (extension to the case of a floating SC terminal can be done \cite{anantram_CurrentFluctuationsMesoscopic_1996}). The orange element denotes a metallic gate whose voltage $V_g$ is tuned to locally deplete the underlying electron gas, reducing the filling factor to $\nu = 1$. As a consequence, edge states with spin parallel to $\vec{B}$ are transmitted, while those with antiparallel spin are reflected. This configuration ensures that the proximized region is reached by edge states originating from different contacts, each kept at a different temperature~\cite{braggio_NonlocalThermoelectricDetection_2024}.

Transport in hybrid QH-SC systems has been widely studied~\cite{giazotto_AndreevReflectionCyclotron_2005,khaymovich_AndreevTransportTwodimensional_2010,vanostaay_SpintripletSupercurrentCarried_2011,sekera_SpinTransportGraphene_2018,beconcini_NonlocalSuperconductingCorrelations_2018,galambos_CrossedAndreevReflection_2022,manesco_MechanismsAndreevReflection_2022,david_GeometricalEffectsDownstream_2023,kurilovich_DisorderenabledAndreevReflection_2023,blasi_TopologicalJosephsonJunctions_2023,kurilovich_CriticalityCrossedAndreev_2023,michelsen_SupercurrentenabledAndreevReflection_2023,arrachea_SignaturesTripletSuperconductivity_2024,maji_SpinOrbitalMixing_2026}.
Here we model the transport properties of our system, mediated by the spin-resolved edge states, using the Landauer-Büttiker approach \cite{datta_ElectronicTransportMesoscopic_1995}. We will always assume to have a non-zero singlet gap $\Delta$ and that $k_B T \ll \Delta$ so that no quasiparticle states are available in the superconductor.
The system's TE properties in the multiterminal setup~\cite{mazza_ThermoelectricEfficiencyThreeterminal_2014} of \Cref{fig:dev} are then quantified by the experimentally accessible Seebeck coefficient $S=\delta V_2/\delta T$, defined as the ratio between the open circuit voltage at contact $2$, $\delta V_2 = (\mu_2-\mu_s)/e$, induced by a thermal bias $\delta T$ in contact $1$.
The explicit form of $S$ is:
\begin{equation}
        S = \frac{\alpha_{21}}{G_{21}+G_{22}} .
    \label{Eq:S}
\end{equation}
Here $G_{ij}={\partial J^{\rm{C}}_i}/{\partial V_j}$ and $\alpha_{ij}={\partial J^{\rm{C}}_i}/{\partial T_j}$
(with $i,j=1,2$)
are, respectively, the linear regime (i.e. when $e\delta V_i,k_B\delta T \ll k_B T$)
electrical conductances and TE coefficients of the currents $J^{\rm{C}}_i$~\cite{lambert_PhasecoherentTransportHybrid_1998,blanter_ShotNoiseMesoscopic_2000,jacquod_OnsagerRelationsCoupled_2012,benenti_FundamentalAspectsSteadystate_2017}.
In terms of the scattering probabilities of the system one finds \cite{supplemental}
\begin{equation}
    \begin{aligned}
        \alpha_{21} &= 
            \frac{2 G_0}{e T}\! \int \! dE
            E
            \left[-f'(E)\right]
            P^A_{21}(E),    \\
        G_{2i} &= 
            (-1)^i G_0 + 2 G_0 \! \int \! dE
            \left[-f'(E)\right]
            P^A_{2i}(E) ,
    \end{aligned}
\label{eq:coeffs}
\end{equation}
where $P^A_{2i}(E)$ is the Andreev transmission probability from contact $i$ to contact $2$, i.e. the probability that an electron with energy $E$ incident from contact $i$ is elastically converted into a hole exiting through contact 2.
We have also defined the conductance quantum $G_0=e^2/h$ and the equilibrium Fermi distribution at temperature $T$, $f(E)=\left[ 1 + \exp( E / k_B T )\right]^{-1}$.

Remarkably, \Cref{eq:coeffs} shows that the TE response vanishes in the absence of superconductivity, since a non-zero $\alpha_{21}$ requires a finite Andreev transmission probability.
Furthermore, for $\alpha_{21}$ to be finite, the Andreev transmission probability $P^{A}_{21}(E)$ must also have an odd-in-energy component.
Indeed, since $E f'(E)$ is an odd function of energy, the corresponding integral vanishes unless $P^{A}_{21}(E)$ breaks EIS, i.e. $P^{A}_{21}(E) \neq P^{A}_{21}(-E)$\cite{arrachea_ThermoelectricProcessesQuantum_2025}.
Normal interchannel electron tunneling between polarized edge states alone is not sufficient to induce a nonvanishing $\alpha_{12}$, but SC ordering is always needed.
Indeed, in our setup, the two edge states after interacting with the superconductor are collected in the same terminal $2$
and any interchannel normal electron tunneling process cannot contribute to TE effects~\cite{supplemental}.

The probabilities $P^A_{2i}(E)$ entering \Cref{eq:coeffs} are entirely determined by the scattering properties of the SC region with length $L$; in all other regions of the system, the edge states propagate freely, merely accumulating phases that do not affect scattering probabilities and are therefore irrelevant.
We model the proximized region using an effective one-dimensional low-energy description: the most general particle-hole-symmetric (Bogoliubov–de Gennes) Hamiltonian for the two edge states, retaining terms up to lowest order in momentum $p$.
Further, we assume that: (i) all SC couplings parameters are constant and nonvanishing only within the proximized region of length $L$ along the transport direction; (ii) the magnetic field is oriented along the $\hat{z}$ direction; and (iii) the superconductor is grounded ($\mu_s=0$). Under these assumptions, the effective Hamiltonian takes the general form:
\begin{equation}
    \begin{aligned}
        \mathbf{H} = \;
        & v p \tau_0\sigma_0
            - B
                \tau_0\sigma_z
            + \vec{v}_s p \cdot \tau_z\vec{\sigma}
        + \Delta \cos(\psi) \tau_x\sigma_0 \\
        & - \Delta \sin(\psi) \tau_y\sigma_0
        + \Re[\vec{v}_\Delta] p \cdot \tau_x\vec{\sigma}
        - \Im[\vec{v}_\Delta] p \cdot \tau_y\vec{\sigma} ,
    \end{aligned}
    \label{eq:ham}
\end{equation}
where $\tau_k\sigma_k$ is the tensor product of Nambu ($\tau_k$) and spin ($\sigma_k$) Pauli matrices, $v$ is the Fermi velocity and $B$ is the Zeeman energy.
We have further introduced the spin-orbit coupling vector $\vec{v}_s$, the singlet order parameter $\Delta e^{i\psi}$, and the complex vector of the triplet coupling $\vec{v}_\Delta$.
The Hamiltonian is written in the basis $(c_{\uparrow},c_{\downarrow},c^\dagger_{\downarrow},-c^\dagger_{\uparrow})$, where the $c_{\gamma}$ ($c^{\dagger}_{\gamma}$) are the destruction operators of the electron-like (hole-like) quasiparticles with spin $\gamma$ on the QH edge states.
It necessarily satisfies particle-hole symmetry by construction, $\mathbf{H} = - \tau_y\sigma_y \mathbf{H}^* \tau_y\sigma_y$ \cite{jacquod_OnsagerRelationsCoupled_2012}.

Further simplifications can be done to obtain a minimal number of parameters that still capture the TE effects induced by the superconductor.
First, we note that spin–orbit coupling can be gauged out into an effective triplet pairing component through an appropriate change of basis when singlet pairing is present~\cite{vanostaay_SpintripletSupercurrentCarried_2011,arrachea_SignaturesTripletSuperconductivity_2024}. 
Therefore, we expect both to have similar physical implications, and we decide then to neglect the spin-orbit term proportional to $\vec{v}_s$ without much loss of generality. 
Further, we assume that the triplet pairing is unitary~\cite{sigrist_IntroductionUnconventionalSuperconductivity_2005}, i.e. $\vec{v}_\Delta \times \vec{v}_\Delta^*=0$ (we have also verified that none of the presented results are crucially affected by this~\cite{supplemental}).
However, this assumption implies that the triplet component has a single SC phase for all spatial components, which can, then, be gauged over the singlet component, which is then characterized by the gauge invariant phase $\psi$ representing the relative phase between triplet and singlet SC components. The Hamiltonian is then simplified to one with a real triplet vector of magnitude $v_\Delta$:
\begin{equation}
    \begin{aligned}
        \mathbf{H} = \;
        & v p \tau_0\sigma_0
            - B \tau_0\sigma_z
            + \Delta \cos(\psi) \tau_x\sigma_0 \\
        & - \Delta \sin(\psi) \tau_y\sigma_0 + v_\Delta p \hat{v}_\Delta \cdot \tau_x\vec{\sigma} .
    \end{aligned}
    \label{eq:hamsim}
\end{equation}
From now on, we will also consider the Fermi velocity $v$ and the gap $\Delta$ to be fixed, hence we will use the normalized quantities  $\tilde{v}_\Delta= v_\Delta/v$ and $\tilde{B}= B/\Delta$.

Starting from the Hamiltonian~(\ref{eq:hamsim}), we can now construct the transfer matrix $\mathbf{M}$ that describes transport across the SC proximized region~\cite{vanostaay_SpintripletSupercurrentCarried_2011,arrachea_SignaturesTripletSuperconductivity_2024}
\begin{equation}
    \begin{gathered}
        \mathbf{M}(\tilde{E})=\exp[i\pi\tilde{L}\left(
            \tilde{E} \mathbf{K}^{-1} - \mathbf{K}^{-1/2} \mathbf{R} \mathbf{K}^{-1/2}
        \right) ],
    \end{gathered}
    \label{eq:transfer}
\end{equation}
where $\tilde{E}=E/\Delta$, $\mathbf{K}=v^{-1}\partial_p \mathbf{H}$ is the normalized velocity matrix, $\mathbf{R} = \left(\mathbf{H}-v\mathbf{K}p\right)/\Delta$ is the $p$-independent matrix, and  $\tilde{L}= 2L \Delta / hv$ is the rescaled length~\footnote{In this effective formalism $hv/2\Delta$ takes the role of a SC coherence length but, in a more microscopical approach, it is also connected with small filling factor changes in the quantum Hall plateau~\cite{david_GeometricalEffectsDownstream_2023}.}.
From \Cref{eq:transfer}, and using the standard relation between scattering and transfer matrices~\cite{datta_ElectronicTransportMesoscopic_1995}, we can finally compute the desired Andreev scattering probabilities $P^A_{2i}(E)$ \cite{supplemental}.
As a final remark, we notice that, due to the geometry of the system and its spin polarization, the scattering properties are invariant under spin rotations around the $\hat {z} $ axis \cite{supplemental}.
Therefore, in the following, we will only consider the angular dependence of TE effects on the zenith angle $\theta$ between the $z$-axis and the triplet vector $ \vec {v}_\Delta$.

\paragraph*{Results.}
The first notable result follows from a direct inspection of \Cref{eq:transfer}: for the system to exhibit TE effects, the matrix $\mathbf{K}$ must not be proportional to the identity matrix. This condition is fulfilled only in the presence of a finite triplet coupling.
In contrast, if triplet pairing is absent, $\mathbf{K}^{-1}$ commutes with all remaining terms in the exponent of the matrix exponential, and the energy $E$ enters only through a global dynamical phase, which cancels out in the scattering probabilities.
Indeed, in the presence of a finite triplet coupling, one generally finds that $\mathbf{M}(\tilde{E})\neq\mathbf{M}(-\tilde{E})$ (except for finely-tuned parameter values). 
Consequently, EIS is broken at the level of Andreev-scattering probabilities, thereby allowing for a finite TE response.

In what follows, we check analytically the validity of this result in the specific case $\theta=0$ and $\psi=0$,
where
the Andreev scattering probabilities
read
\begin{equation}
    P^A_{2i}(E) = \sin^2\left(
        \pi\tilde{L}
        \left[
            \frac{
                1 + \tilde{v}_\Delta (
                    \tilde{B} - (-1)^i \tilde{E}
                )
            }{
                1 - \tilde{v}_\Delta^2
            }
        \right]
    \right).
    \label{eq:probs}
\end{equation}

\begin{figure}
    \includegraphics[width=\linewidth]{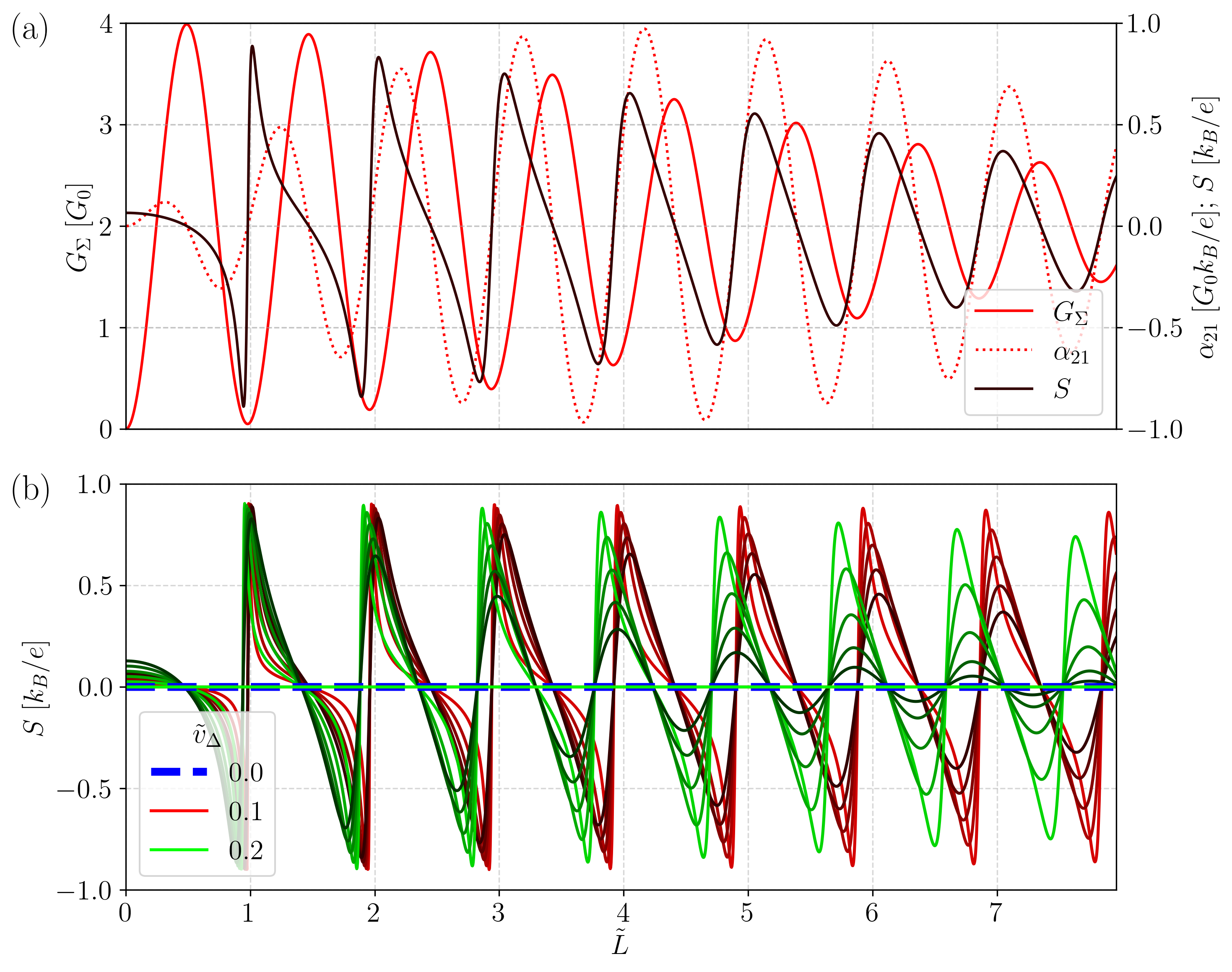}
    \caption{Transport coefficients as functions of length $\tilde{L}$ for $\theta=\psi=0$ and $\tilde{B}=0.1$. Panel (a): conductance $G_{\Sigma}$ (solid red curve), $\alpha_{21}$ (red dotted curve), and $S$ (solid black curve), each shown with its own scale, computed at $\tilde{T}=0.2$ and $\tilde{v}_\Delta=0.1$. Panel (b): Seebeck coefficient $S$ for different values of $\tilde{v}_\Delta$ ($0.0$, $0.1$, and $0.2$, shown in blue, red, and green, respectively) and different values of $\tilde{T}$ (ranging from $0$ in light shades to $0.2$ in darker shades). For $\tilde{v}_\Delta=0$, $S$ vanishes identically, and the corresponding thick dashed blue curve lies along the horizontal axis.
    }
    \label{fig:seeb-vdelta-temp}
\end{figure}

Using Eqs.~(\ref{Eq:S}) and (\ref{eq:coeffs}) we can compute the Seebeck coefficient, obtaining
\begin{equation}
    \begin{aligned}
        S &= \frac{k_B}{e}
        \frac{
            \pi \sin(\lambda) \csch(\eta)
            \left[
                \eta \coth(\eta) - 1
            \right]
        }{
            2 \left[
                1 - \cos(\lambda) \eta \csch(\eta)
            \right]
        },
    \end{aligned}
    \label{eq:See}
\end{equation}
where
\begin{equation}
    \eta = 2\pi^2 \frac{
        \tilde{v}_\Delta
    }{
        1-\tilde{v}_\Delta^2
    } \tilde{L} \tilde{T} , ~~~~
    \lambda = 2\pi \frac{
        1 + \tilde{v}_\Delta \tilde{B}
    }{
        1-\tilde{v}_\Delta^2
    } \tilde{L} ,
        \label{etalambda}
\end{equation}
with $\tilde{T}= k_B T/\Delta$ being the normalized temperature.
One can immediately observe that in the absence of triplet pairing, i.e.~$\tilde{v}_\Delta \rightarrow 0$, the thermopower $S$ vanishes, thereby highlighting the crucial role of the triplet component in generating the TE effect.

In \Cref{fig:seeb-vdelta-temp}(a) we illustrate the behavior of $S$, its numerator $\alpha_{21}$ (right scale) and denominator $G_\Sigma=G_{21}+G_{22}$ (left scale), as functions of the normalized length $\tilde{L}$.
Panel~(a) shows that $\alpha_{21}$ (red dotted line) and the thermopower $S$ (black solid curve) oscillate around zero, while $G_{\Sigma}$ (red solid line) oscillates about $2G_0$ while remaining strictly positive, as expected from the analytical expressions in \Cref{eq:coeffs}.
As evidenced by \Cref{eq:See}, oscillations are controlled by the parameter $\lambda$ (which is a function of $\tilde v_\Delta$, $\tilde B$, and $\tilde L$), and originate from interference effects arising from the interplay between crossed Andreev reflection and normal transmission along the edge~\cite{khaymovich_AndreevTransportTwodimensional_2010,david_GeometricalEffectsDownstream_2023}.
The thermopower $S$, in addition, exhibits a peculiar resonant-like structure near small integer values of $\tilde L$, which coincide with the zeros of the TE coefficient $\alpha_{21}$. This behavior originates from the fact that these values of $\tilde L$ also correspond to local (quadratic in $\tilde L$) minima of $G_\Sigma$. Since $\alpha_{21}$ vanishes only linearly with $\tilde L$, the resulting ratio gives rise to sharp, sign-changing peaks in $S$~\footnote{Notably, in such a system, maximizing the thermoelectric coefficient $\alpha_{21}$ does not correspond to maximizing the total Andreev processes, but rather their component that is odd in energy.}.

On the other hand, for large values of $\tilde{L}$ \cite{supplemental}, both $S$ and $\alpha_{21}$ are suppressed to zero, while the oscillations of the conductance $G_{\Sigma}$ are damped towards $2G_0$. This behavior stems from the energy dependence of the Andreev probability $P^A_{21}(E)$ (in \Cref{eq:probs}) induced by the triplet pairing, which also depends on $\tilde{L}$. As $\tilde{L}$ increases, the period of the oscillations in energy of $P^A_{21}(E)$ decreases as  
$\tilde{L}^{-1}$, so that they are progressively averaged out over the thermal energy window, leading to a suppression of the oscillations in $\tilde{L}$~\cite{supplemental}.

Panel~(b) of \Cref{fig:seeb-vdelta-temp} shows the length-dependence of the Seebeck coefficient $S$ for different values of the triplet pairing parameter $\tilde v_\Delta$ and temperature $\tilde T$.
We start by emphasizing that $S$ vanishes identically when $\tilde v_\Delta = 0$ (blue thick dashed line). By contrast, at finite temperature, a nonzero thermopower arises whenever triplet pairing is present, as illustrated by the two cases considered, $\tilde v_\Delta = 0.1$ (red curves) and $\tilde v_\Delta = 0.2$ (green curves).
Focusing then on the temperature dependence, in the plot $\tilde{T}$ ranges from $0$ to $0.2$, with increasing temperature indicated by a progressive darkening of the color (from bright to dark). We first consider the green ($\tilde v_\Delta = 0.2$) curves. As a general trend, increasing the temperature leads to a reduction in the amplitude of $S$, as a consequence of thermal averaging over a broader energy (thermal) window. This suppression is more pronounced at larger values of $\tilde{L}$, where the averaging becomes more effective, resulting in a damping of the oscillations with $\tilde{L}$ at fixed temperature $\tilde{T}$ (the damping becoming stronger as $\tilde{T}$ increases).
A similar behavior is observed for the red ($\tilde v_\Delta=0.1$) curves, although the effect of thermal averaging is less pronounced, resulting in weaker damping. Overall, the damping length decreases with increasing temperature and with increasing triplet coupling. This is because $\tilde T$ and $\tilde v_\Delta$ have a similar effect on the parameter $\eta$ (see \Cref{etalambda}) that controls the envelope of the thermopower oscillations in \Cref{eq:See}, leading to a suppression of $S$ as soon as $\eta \gtrsim 1.5$~\cite{supplemental}.
Finally, we note that $\tilde v_\Delta$ also enters the parameter $\lambda$, so that the strength of the triplet pairing slightly modifies the periodicity of $S$ in $\tilde L$, as evidenced by the increasing horizontal shift between the red and green curves.

\begin{figure}
    \includegraphics[width=\linewidth]{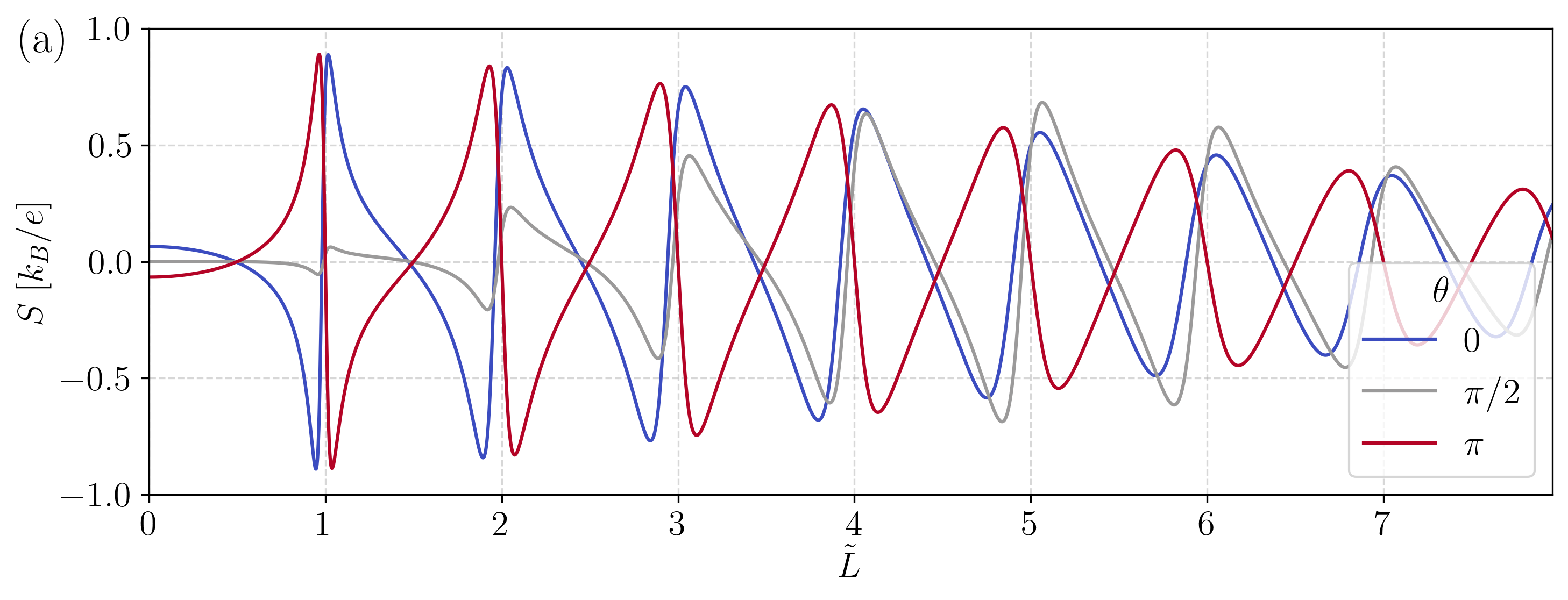}
    \includegraphics[width=\linewidth]{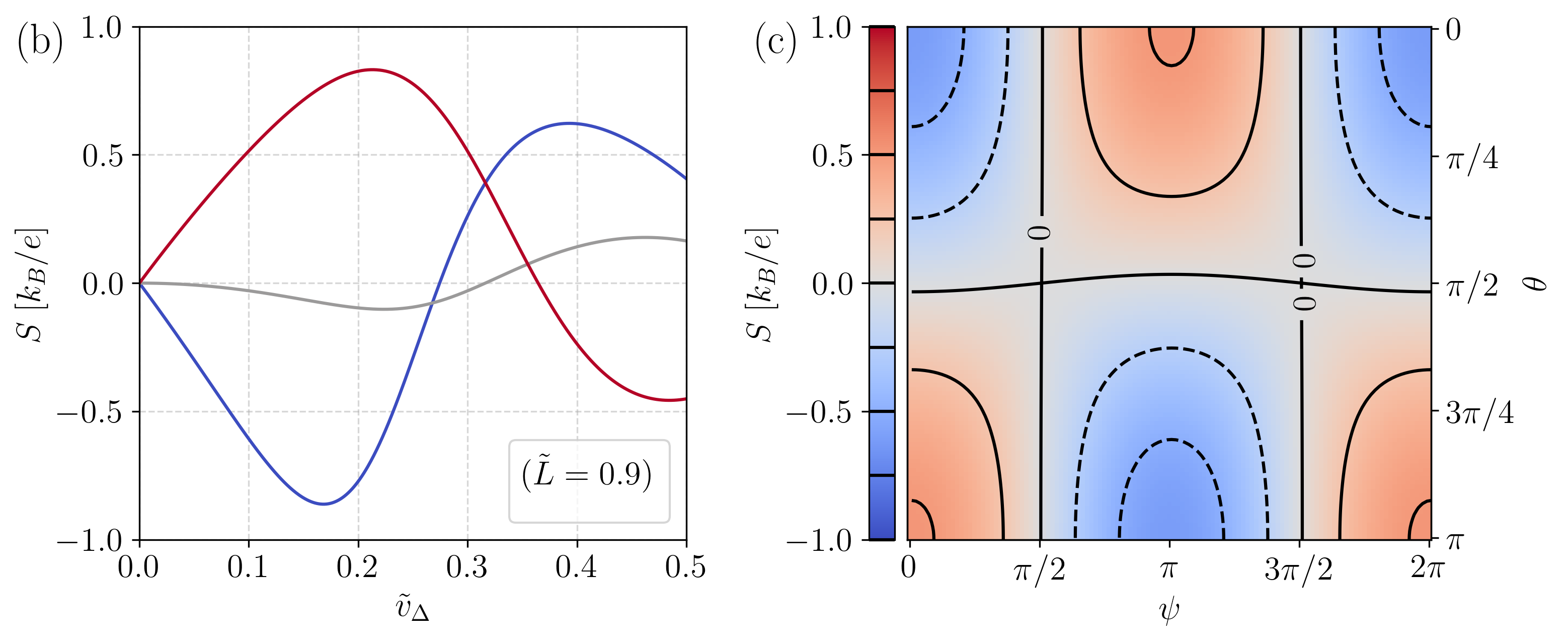}
    \caption{Behavior of the thermopower $S$ for finite values of $\theta$ and $\psi$ at $\tilde{T}=0.2$, $\tilde{B}=0.1$:
    (a) as a function of $\tilde{L}$ for different values of $\theta$, at $\psi=0$ and $\tilde{v}_\Delta=0.1$.
    (b) as a function of $\tilde{v}_\Delta$, for the same values of $\theta$ in panel (a), at $\psi=0$ and $\tilde{L}=0.9$.
    (c) as a density plot in the $\psi$-$\theta$ space, at $\tilde{v}_\Delta=0.1$ and $\tilde{L}=0.9$.
    }
    \label{fig:theta}
\end{figure}

The above analysis of Fig.~\ref{fig:seeb-vdelta-temp} remains qualitatively valid even when the assumptions $\theta=\psi=0$ are relaxed, although a richer phenomenology emerges. In \Cref{fig:theta}, starting from panel (a), we observe how the thermopower evolves as a function of $\tilde{L}$ for different values of the azimuthal angle $\theta$. 
In particular, when the triplet vector is parallel or antiparallel to the magnetic field, i.e., $\theta=0$ or $\theta=\pi$, the Seebeck coefficient $S$ (shown by the blue and red curves, respectively) approximately reverses sign between the two configurations. This behavior reflects the fact that inverting the triplet vector effectively exchanges the roles of the two spin channels, which is physically equivalent to reversing the temperature gradient between them.
The symmetry under spin reversal, however, is not exact, so that the $\theta=0$ and $\theta=\pi$ cases are not perfectly mirrored about the $S=0$ line, due to the presence of the Zeeman term in the Hamiltonian~(\ref{eq:hamsim}), which also slightly modifies the oscillation frequency of $S$ between the two cases. In the absence of Zeeman coupling, which is hard to be practically realized in our setup, this mirror symmetry is fully restored~\cite{supplemental}.
For intermediate angles, such as $\theta=\pi/2$ (grey curve), $S$ displays instead a beating pattern, arising from interference processes, over a broad range of $\tilde{L}$. \cite{supplemental}.

Figure~\ref{fig:theta}(b) shows the oscillatory dependence of the Seebeck coefficient on the triplet coupling strength $\tilde v_\Delta$, evaluated at fixed $\tilde{L}$ and $\psi=0$, for three representative values of $\theta$. This behavior can be attributed to the fact that varying $\tilde{v}_\Delta$ produces a shift in the position of the Seebeck resonance, analogous to the shift observed in \Cref{fig:seeb-vdelta-temp}(b) when comparing the green and red curves.
In panel (c), we finally present the dependence of the thermopower at fixed $\tilde{L}$ and $\tilde{v}_\Delta$ in the $\theta$-$\psi$ plane.
First of all, $S$ exhibits periodic oscillations in $\psi$: for fixed $\theta$, the behavior is almost cosinusoidal, with the zeroes located at $\pi/2$ and $3\pi/2$. This suggests that the thermopower can serve as a potential probe of the phase difference between the singlet and the triplet SC components.
In $\theta$ instead, $S$ changes sign at approximately $\theta=\pi/2$; if there was no Zeeman splitting, this would occur exactly at $\pi/2$~\cite{supplemental}.

\paragraph*{Conclusions.}
In this work, we have shown that a QH bar proximized by a superconductor can display a finite linear thermoelectric response in the middle of the QH plateau, with a thermopower $S$ reaching values on the order of $k_B/e$.
Surprisingly, this low-energy effect arises in the subgap regime and in the presence of a linear dispersion of the chiral edge states, two conditions that, when considered separately, would normally imply no thermoelectricity in itself.
In our setup, on the contrary, the TE effect is allowed thanks to the triplet component of the SC pairing, since it
introduces momentum-dependent particle-hole mixing terms and, consequently, breaks the EIS of the scattering probabilities between the two edge states when they are spin-polarized. 
Further, we found that $S$ oscillates as a function of the length of the proximized superconducting region and of the magnitude and orientation of the triplet pairing vector. The thermopower is also sensitive to the singlet-triplet relative phase.

We expect our minimal model to hold around the middle of the QH plateau.
In fact, from the oscillatory behavior in $\tilde L$ one can infer a corresponding oscillatory dependence also on other tunable experimental parameters, such as variations of the filling factor within the plateau, achievable via a back gate~\cite{david_GeometricalEffectsDownstream_2023}.
Oscillatory behavior of the TE response has been recently observed  \cite{mccourt_ThermoelectricEffectQuantum_2025,wang_ThermoelectricDetectionCrossed_2026}, albeit it has been interpreted in terms of fluxon dynamics.
Considering other aspects, such as the inclusion of inter-edge Coulomb interactions, they are expected to produce only quantitative, but not qualitative, modifications of the predicted oscillations~\cite{braggio_NonlocalThermoelectricDetection_2024}.
We hope that our results will further stimulate experimental investigations of linear TE effects in QH-SC hybrid systems.

\paragraph*{Acknowledgments.} We thank P. Burset and L. Arrachea for fruitful interaction.
We acknowledge funding from MUR-PRIN 2022 - Grant No. 2022B9P8LN
- (PE3)-Project NEThEQS “Non-equilibrium coherent
thermal effects in quantum systems” in PNRR Mission 4 -
Component 2 - Investment 1.1 “Fondo per il Programma
Nazionale di Ricerca e Progetti di Rilevante Interesse
Nazionale (PRIN)” funded by the European Union - Next
Generation EU and the PNRR MUR project
PE0000023-NQSTI.
AB acknowledges the project "Thermoelectric effects in solid-state quantum devices based on multiterminal Josephson junctions” of the bilateral agreement CNR/CONICET (Italy/Argentina) 2026-2027 and the CNR Project QTHERMONANO.

\onecolumngrid

\appendix

\section{Scattering matrix and probabilities}
In this section we present the detailed form of the scattering matrix and the subsequent derivation of the relevant scattering probabilities and transport coefficients of the main text.
The system under consideration is defined in the manuscript, and it describes the situation represented in Fig. 1 of the main text.

Considering that there are two normal contacts with two open transport channels each (for spin-up and spin-down electrons), and that in the subgap energy regime there are no scattering states propagating in the superconductor, the scattering matrix of the system is an $8\times 8$ matrix,  also accounting for the quasiparticle/quasihole degree of freedom.
Given the geometry of the QH bar, there are no backreflections of scattering states unless dictated by the gating at $V_g$; if the edge state for a particle of type $\alpha$ ($e$ for quasi-particles and $h$ for quasi-holes) with spin $\sigma$ does not mix with any other during its travel from contact $j$ to $i$, it just accumulates an irrelevant phase $\varphi^{\alpha}_{ij\sigma}(E)$.
In addition, for simplicity, hereafter we assume to neglect any 
normal electron tunneling (spin mixing) between the edge states when we discuss the scattering matrix. We shall see that those processes cannot contribute to the TE properties in the proposed setup anyway.

The scattering matrix in the basis $(c_{1\uparrow},c_{1\downarrow},c_{2\uparrow},c_{2\downarrow},c^\dagger_{1\uparrow},c^\dagger_{1\downarrow},c^\dagger_{2\uparrow},c^\dagger_{2\downarrow})^T$ takes the form (we implicitly assume the energy dependence of the matrix elements):
\begin{equation}
    \mathbf{s}(E)=
    \begin{NiceMatrixBlock}
        \begin{pNiceArray}%
        {%
            c:c|%
            c:c%
        }
            s^{ee}_{11} &   s^{ee}_{12} &   s^{eh}_{11} &   s^{eh}_{12} \\
            \hdottedline
            s^{ee}_{21} &   s^{ee}_{22} &   s^{eh}_{21} &   s^{eh}_{22} \\
            \hline
            s^{he}_{11} &   s^{he}_{12} &   s^{hh}_{11} &   s^{hh}_{12} \\
            \hdottedline
            s^{he}_{21} &   s^{he}_{22} &   s^{hh}_{21} &   s^{hh}_{22} \\
        \end{pNiceArray}
    \end{NiceMatrixBlock} =
    \begin{NiceMatrixBlock}
        \begin{pNiceArray}%
        {%
            cc:cc|%
            cc:cc%
        }
            0   &
                0   &
                e^{
                    i\varphi^{e}_{
                        2 1\uparrow
                    }
                }   &
                0   &
                0   &
                0   &
                0   &
                0   \\
            0   &
                e^{
                    i\varphi^{e}_{
                        1 1\downarrow
                    }
                }   &
                0   &
                0   &
                0   &
                0   &
                0   &
                0   \\
            \hdottedline
            m_{e\uparrow e\uparrow}  &
                0   &
                0   &
                m_{e\uparrow e\downarrow}  &
                m_{e\uparrow h\uparrow}  &
                0   &
                0   &
                m_{e\uparrow h\downarrow}  \\
            m_{e\downarrow e\uparrow}  &
                0   &
                0   &
                m_{e\downarrow e\downarrow}  &
                m_{e\downarrow h\uparrow}  &
                0   &
                0   &
                m_{e\downarrow h\downarrow}  \\
            \hline
            0   &
                0   &
                0   &
                0   &
                0   &
                0   &
                e^{
                    i\varphi^{h}_{
                        2 1\uparrow
                    }
                }   &
                0   \\
            0   &
                0   &
                0   &
                0   &
                0   &
                e^{
                    i\varphi^{h}_{
                        1 1\downarrow
                    }
                }   &
                0   &
                0   \\
            \hdottedline
            m_{h\uparrow e\uparrow}  &
                0   &
                0   &
                m_{h\uparrow e\downarrow}  &
                m_{h\uparrow h\uparrow}  &
                0   &
                0   &
                m_{h\uparrow h\downarrow}  \\
            m_{h\downarrow e\uparrow}  &
                0   &
                0   &
                m_{h\downarrow e\downarrow}  &
                m_{h\downarrow h\uparrow}  &
                0   &
                0   &
                m_{h\downarrow h\downarrow}  \\
        \end{pNiceArray}
    \end{NiceMatrixBlock},
    \label{eq:scatt}
\end{equation}
where $m_{\alpha \sigma\beta\sigma'}$ are the elements of the transfer matrix $\mathbf{M}(\Tilde{E})$ defined in the main text, although now defined over the basis $(c_\uparrow,c_\downarrow,c^\dagger_\uparrow,c^\dagger_\downarrow)^T$.
In general, the scattering probabilities are defined as:
\begin{equation}
    P_{ij}(E)
        = \Tr[s^{ee\dagger}_{ij}(E)s^{ee}_{ij}(E)]
        = \Tr[s^{hh\dagger}_{ij}(-E)s^{hh}_{ij}(-E)]; \quad
    P^A_{ij}(E)
        = \Tr[s^{he\dagger}_{ij}(E)s^{he}_{ij}(E)]
        = \Tr[s^{eh\dagger}_{ij}(-E)s^{eh}_{ij}(-E)],
    \label{eq:phsprob}
\end{equation}
where the second equalities are necessarily valid due to PHS.
According to the Landauer-Büttiker scattering theory~\cite{blanter_ShotNoiseMesoscopic_2000,lambert_PhasecoherentTransportHybrid_1998}, the electrical currents in the linear regime are defined as:
\begin{equation}
    J^{\rm{C}}_i = \frac{e}{h} \int_{-\infty}^{+\infty} dE
    \left[-f'(E)\right]
    \sum_{j=1}^2\left[
        N_i(E)\delta_{ij} - P_{ij}(E) + P^A_{ij}(E)
    \right]
    \left( e \delta V_j + E \delta T_j/T \right),
\end{equation}
with $f(E)=\left[ 1 + \exp(E/k_B T) \right]^{-1}$ and $N_i(E) \equiv 2\; \forall i$ in our case.
Consequently, the electrical conductances $G_{ij}$ and TE coefficients $\alpha_{ij}$ can be easily calculated. In particular, for our setup, by using the unitarity of the scattering matrix $\sum_{i=1}^2 P_{ij}(E)+P_{ij}^A(E) = N_j(E), \forall E$, we can write down the scattering probabilities:
\begin{equation}
    \mathbf{P}(E)=
    \begin{NiceMatrixBlock}
        \begin{pNiceArray}%
        {%
            c:c|%
            c:c%
        }
            P_{11}(E)   &   P_{12}(E)   &   P^A_{11}(-E)    &   P^A_{12}(-E)    \\
            \hdottedline
            P_{21}(E)   &   P_{22}(E)   &   P^A_{21}(-E)    &   P^A_{22}(-E)    \\
            \hline
            P^A_{11}(E) &   P^A_{12}(E) &   P_{11}(-E)  &   P_{12}(-E)  \\
            \hdottedline
            P^A_{21}(E) &   P^A_{22}(E) &   P_{21}(-E)  &   P_{22}(-E)
        \end{pNiceArray}
    \end{NiceMatrixBlock}
    = \begin{NiceMatrixBlock}
        \begin{pNiceArray}%
        {%
            c:c|%
            c:c%
        }
            1   &   1   &   0   &   0   \\
            \hdottedline
            1-P^A_{21}(E)   &   1-P^A_{22}(E)   &   P^A_{21}(-E)    &   P^A_{22}(-E)    \\
            \hline
            0   &   0   &   1   &   1   \\
            \hdottedline
            P^A_{21}(E) &   P^A_{22}(E) &   1-P^A_{21}(-E)  &   1-P^A_{22}(-E)
        \end{pNiceArray}
    \end{NiceMatrixBlock},
    \label{eq:scattprobs}
\end{equation}
which takes this simple form, with terms being either 0, 1, or simply dependent on $P^A_{21}(E),P^A_{22}(E)$ due to the chiral nature of the edges.
Therefore, the coefficients of interest for us are:
\begin{equation}
    \begin{aligned}
        G_{21} &= G_0 \int_{-\infty}^{+\infty} dE
            \left[-f'(E)\right]
            \left[
                - P_{21}(E) + P^A_{21}(E)
            \right]
            = G_0 \int_{-\infty}^{+\infty} dE
            \left[-f'(E)\right]
            \left[
                - 1 + 2 P^A_{21}(E)
            \right]; \\
        G_{22} &= G_0 \int_{-\infty}^{+\infty} dE
            \left[-f'(E)\right]
            \left[
                2 - P_{22}(E) + P^A_{22}(E)
            \right]
            = G_0 \int_{-\infty}^{+\infty} dE
            \left[-f'(E)\right]
            \left[
                1 + 2 P^A_{22}(E)
            \right]; \\
        \alpha_{21} &= \frac{e}{h T} \int_{-\infty}^{+\infty} dE
            \left[-f'(E)\right] E
            \left[
                - P_{21}(E) + P^A_{21}(E)
            \right]
            = \frac{2e}{h T} \int_{-\infty}^{+\infty} dE
            \left[-f'(E)\right] E
            P^A_{21}(E); \\
    \end{aligned}
\end{equation}
where we have used the unitarity of the scattering matrix and the fact that the integral on the real axis of $f'(E)E$ is zero.

It is interesting to note that the last expression of $\alpha_{21}$ clarifies that any normal spin-flip tunneling process 
alone cannot induce TE effects in our setup and, instead, SC coupling is strictly necessary. 
Indeed, the proposed setup does not collect separately the current of the two spin channels exiting from the scattering region through separate electrical contacts but, instead, they are all collected at contact 2.
Consequently, electron spin-flip scattering processes between the edge states have no net effect on the total $J_2$ current.
Instead, by Andreev reflections, the superconductor can inject a net charge into the two edge states as soon as a thermal bias is established between the two incoming 
channels before they enter in the scattering area, if a triplet component is present and the two edge channel are polarized.
As a final remark we note that, even if normal spin-flip tunneling processes between the edge states cannot induce thermoelectricity by themselves, they could still impact the strength of the superconductivity-induced thermoelectricity due to the unitarity: for example, if $P_{ij}(E)=1$, then Andreev processes are necessarily suppressed, i.e. $P_{ij}^A(E)=0$.  

\subsection{Invariance under triplet vector rotations around \texorpdfstring{$\hat z$}{z} axis}

As we noted in the main text in the discussion after Eq.~(5), the transport properties of the system do not depend on the orientation of the triplet vector $\vec{v}_\Delta$, which is characterized by its azimuthal and zenith orientation angles $\varphi$ and $\theta$, around the magnetic field direction $\hat z$.
In fact, the triplet azimuthal orientation can be seen as an effective rotation $U_z(-\varphi)$ in spin space.
Then, we would get $\mathbf H \rightarrow U_z(-\varphi)\,\mathbf{H}\,U_z^\dagger(-\varphi)$, with the transfer matrix $\mathbf{M}(\tilde E)$ transforming 
similarly.
Such rotation would result in the scattering amplitudes being multiplied by phase factors; nonetheless, since $P^A_{21}(E) \coloneqq |m_{h\uparrow e\uparrow}|^2 + |m_{h\downarrow e\uparrow}|^2$ (see \Cref{eq:scatt,eq:phsprob}), such phases would cancel out, leaving all transport observables unaffected.

\section{Other results and observations}

In the following, we will describe further details of the dependence of the TE effect in peculiar cases and on other parameters of the Hamiltonian, not explored in detail in the main text.

\subsection{\texorpdfstring{Oscillation damping of the transport coefficients and $\tilde{v}_\Delta=0$}{vΔ} case}

\begin{figure}
    \includegraphics[height=5.0cm]{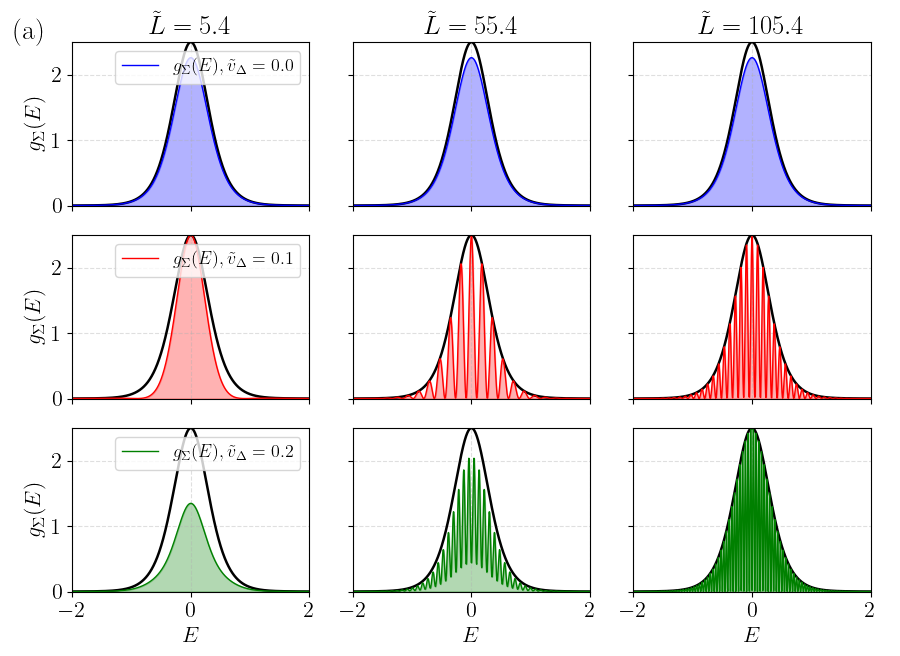}%
    \includegraphics[height=5.0cm]{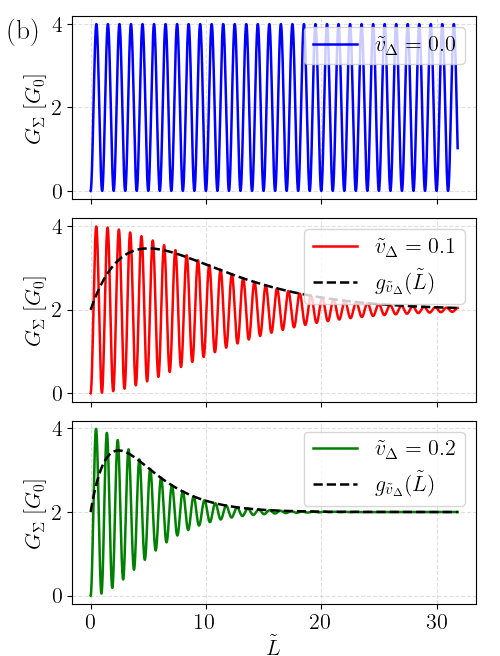}%
    \includegraphics[height=5.0cm]{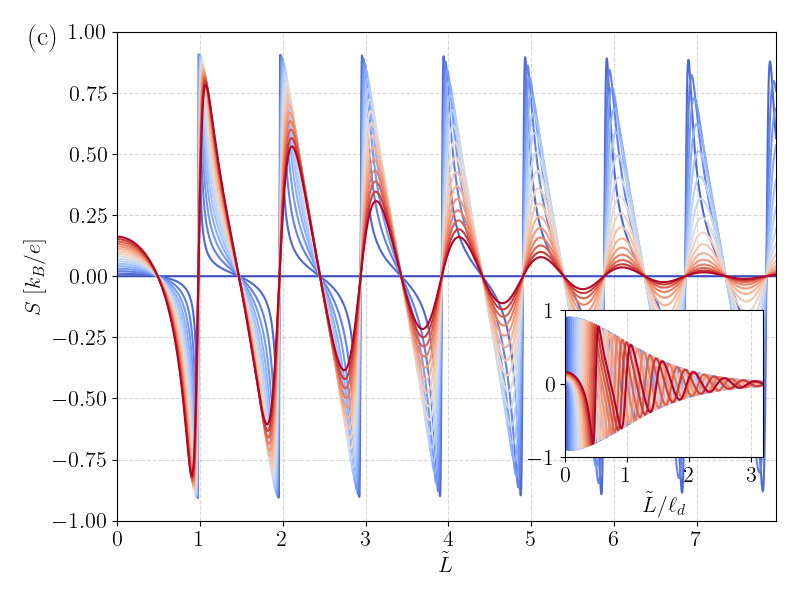}
    \caption{(a): Plots of the functions in the integral defining $G_\Sigma=\int dE g_\Sigma(\tilde E)$, for $\theta,\psi=0$ and $\tilde B=0.1$.
    In black, as a visual guideline, the Fermi function derivative times -2 is plotted, at $\tilde T=0.1$: since $P^A_{21}+P^A_{22}$ has absolute maximum 2, it encompasses fully the region covered by the plot of the $g_\Sigma(E)$.
    in different rows, the colored lines are the energy dependent integrands $g_\Sigma(\tilde E)$, for increasing $\tilde{v}_\Delta$; different columns correspond to different values of $\tilde L$; since the periodicity of $G_\Sigma$ in $\tilde L$ is 1 when $\tilde v_\Delta=0$, the three plots are identical, since the shown values of $\tilde L$ differs by an integer amount.
    When $\tilde{v}_\Delta\neq 0$, the integrand starts to oscillate in energy.
    (b): plots of $G_\Sigma$ as a function of $\tilde{L}$ for very long distances.
    For $\tilde{v}_\Delta=0$ it is exactly periodic, but with triplet $\tilde{v}_\Delta\neq 0$ it damps to the average value $2G_0$ for $\tilde{L}\to\infty$ 
    The black dashed lines, in the $\tilde v_\Delta = 0.1,0.2$ cases, are the asymptotic envelope functions $g_{\tilde v_\Delta}$ discussed in the text.
    (c): Plots of $S$ for different temperatures, from $\tilde T=0.0$ (in blue) to $0.5$ (in red), at $\tilde B=0.1$, $\tilde v_\Delta$ and $\theta,\psi=0$.
    The inset shows the damping of all the plots, but in length units rescaled with the damping length $\ell_d$: all the plots collapse under one single envelope function.
    }
    \label{fig:vd0-integral}
\end{figure}
Notably, according to Eqs. (2) and (6) of the main text, in the limit \(\tilde v_\Delta \to 0\), i.e. only singlet superconductivity, the conductance \(G_\Sigma\) remains finite and oscillates as \([1-\cos(2\pi\tilde L)]\), 
without any damping.
This result is consistent with the expected oscillation of the conductance due to the alternating conversion between electrons and holes, which, for appropriate geometries, yields alternating positive and negative nonlocal conductances.
In this limit, the scattering probabilities \(P^A_{2i}(E)\) in Eq. (6) of the main text become energy-independent but oscillate with \(\tilde L\) with period \(1\)~\cite{khaymovich_AndreevTransportTwodimensional_2010,david_GeometricalEffectsDownstream_2023,beconcini_NonlocalSuperconductingCorrelations_2018}.
This means that the oscillatory behavior of $G_\Sigma$ is independent from the triplet pairing. However, with triplet components, the conductance oscillations damp.

To better clarify this, in panel (a) of \Cref{fig:vd0-integral} we show the behavior of the functions of the integrand $g_\Sigma(E)$ defining $G_\Sigma=\int dE g_\Sigma(\tilde E)$. The actual integral corresponds to the shaded colored area under the colored curve.
We consider the cases corresponding to the triplet values discussed in Fig. 2 of the main text, taken at various $\tilde L$ values (see labels on top) and $\tilde v_\Delta $ (different colors). Those can be compared with the corresponding plots in panel (b) for $G_\Sigma$ computed as a function of the $\tilde L$ length for different values of $v_\Delta $(different colors).

The first row of panel (a) explicitly shows why, for $\tilde {v}_\Delta=0$,  $G_\Sigma$ always oscillates without 
decaying with $~\tilde{L}$, even for very large $\tilde{L}$ (see top (b) panel).
The total sum of Andreev scattering probabilities $P^A_{21}(E)+P^A_{22}(E)$ is constant in energy at fixed $\tilde L$. Consequently, $g_\Sigma$ stays within the profile of $-2f'(E)$ (solid black line, which is the Fermi function derivative times the absolute maximum of $P^A_{21}(E)+P^A_{22}(E)$), but even if its value oscillates in $\tilde L$ it has exactly period 
\(1\).
In panel (a), we show the integrand for different lengths, but all in the same position with respect to the period structure \(1\). 
In the other rows of the panel (a), we instead see why there is damping: when $v_{\Delta}\neq 0$, the sum of the Andreev probabilities $P^A_{21}(E)+P^A_{22}(E)$ shows increasing energy oscillations with increasing $\tilde v_\Delta$ or $\tilde L$. The same happens for the plotted integrand $g_\Sigma (E)$. 
Therefore, the scattering processes at different energies tend to average out inside the thermal window (determined by $-f'(E)$), so that it progressively approximates half the area under $-2f'(E)$, exactly corresponding to the limit
$2G_0$ for $\tilde{L}\to\infty$.

To see how fast $G_\Sigma$ damps out in $\tilde L$, we have shown in panel (b) with a black line the approximated envelope function for the local maxima:
\begin{equation}
    g_{\tilde v_\Delta}(\tilde L) = 2 ( 1 + 2 \tilde L / \ell_d e^{-\tilde L/\ell_d}),
    \label{eq:envelope}
\end{equation}
which is asymptotically valid for $\tilde{L}/\ell_d\to\infty$,
where the damping length is $\ell_d=(1-\tilde v^2_\Delta)/(2\pi^2\tilde T\tilde v^2_\Delta)$.
As we can see, the damping is more than exponential and, as anticipated, depends on the quantities determined by the definition of the variable $\eta$ in the main text.

Finally, in panel (c), the $\tilde{L}$ behavior of the thermopower for different temperatures is shown.
The interplay between the damping length and temperature is clearly shown: because each colored line corresponds to a different temperature (cold is blue, hot is red), the damping length $\ell_d$ shortens with increasing temperatures.
Therefore, in the inset, we show that, if the lengths are rescaled by the damping length, i.e., $\tilde L/\ell_d$, all the curves of the Seebeck coefficient calculated at different temperatures collapse onto the same curve, as expected from Eq.~\ref{eq:envelope}. However, clearly, with such scaling, the number of oscillations varies widely.

\subsection{Angle \texorpdfstring{$\theta$}{θ} and phase \texorpdfstring{$\psi$}{ψ} dependences, with and without Zeeman splitting}

\begin{figure}
    \includegraphics[width=0.5\linewidth]{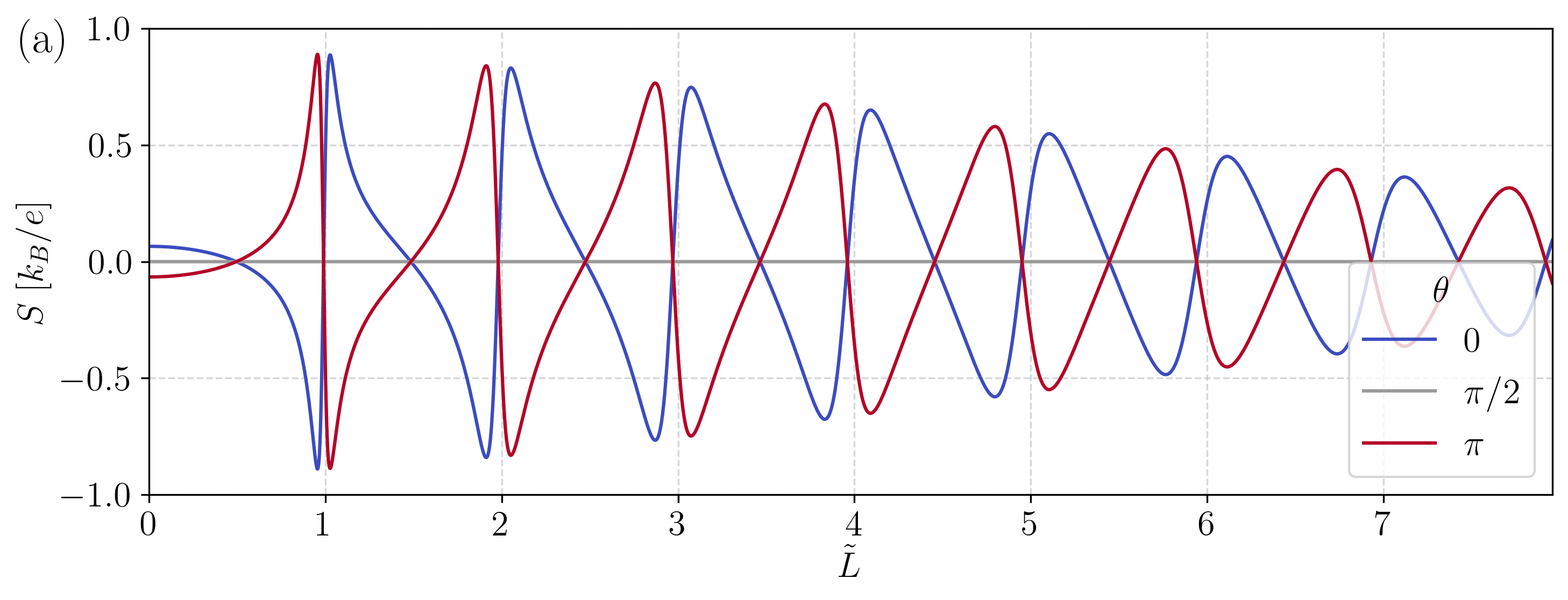}%
    \includegraphics[width=0.5\linewidth]{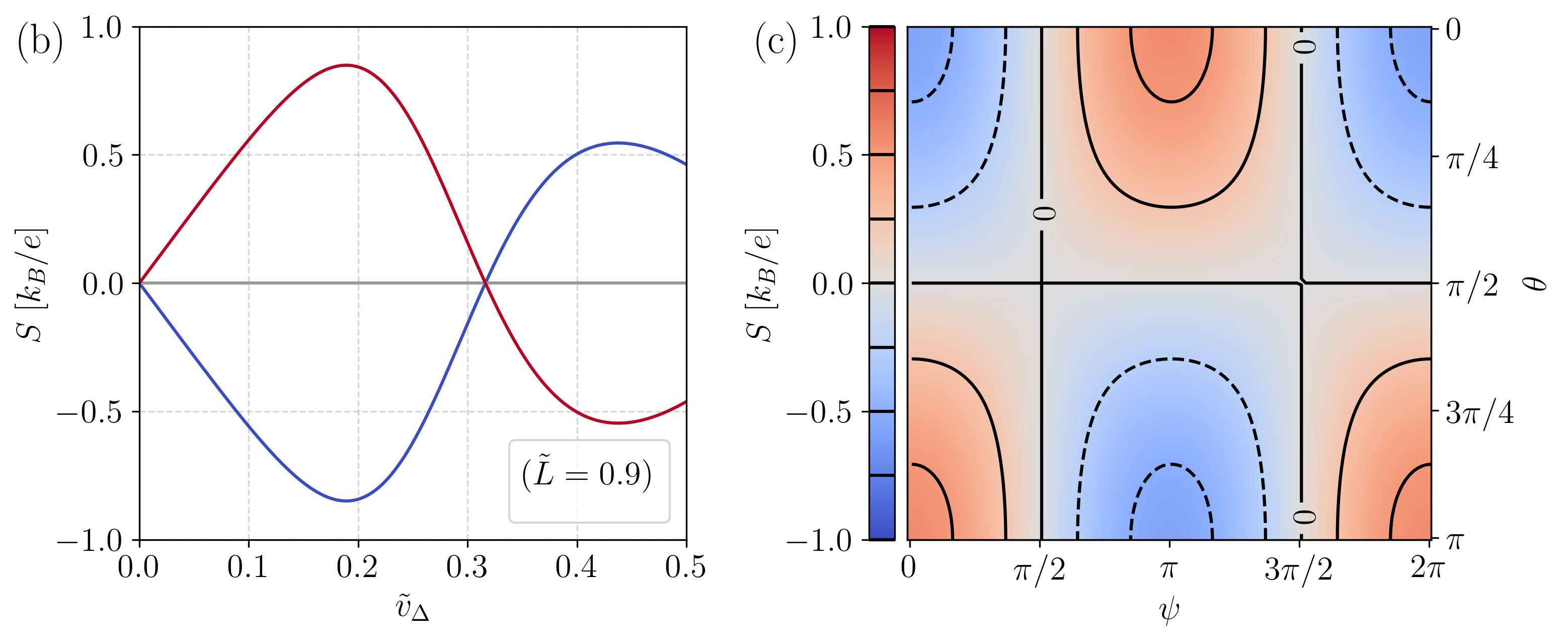}
    \caption{Thermopower dependence on $\tilde{L}$, $\theta$, $\tilde{v}_\Delta$ and in the $\theta-\psi$ plane, with $\tilde{T}=0.1$ and $\tilde{B}=0$.}
    \label{fig:thetab0}
\end{figure}

In \Cref{fig:thetab0}, we show the Seebeck coefficient, as in Fig. 3 of the main text, but for zero Zeeman splitting.
We do not analyze this case in the main text because this limit, while theoretically interesting, cannot be easily realized in an experiment, given that the  Zeeman term is required to have spin-polarized edge states and a strong magnetic field is required in the quantum Hall regime.
However, in the theoretical absence of Zeeman splitting, the TE properties are exactly antisymmetric under the transformation $\theta\rightarrow\pi-\theta$, with the Seebeck coefficient correspondingly being $S\rightarrow -S$.
Clearly, we observe that for $\theta=\pi/2$ the (grey lines in (a) and (b)) the Seebeck coefficient is always zero.
Finally, we also see that, at fixed $\theta$, the dependence in phase $\psi$ appears cosinusoidal and antisymmetric under reflection with respect to $\theta=\pi/2$.
This clearly shows that it is the introduction of the Zeeman term that breaks the perfect symmetries and antisymmetries in the angle and phase dependencies of the figure reported in the main text.

\begin{figure}
    \includegraphics[width=0.45\linewidth]{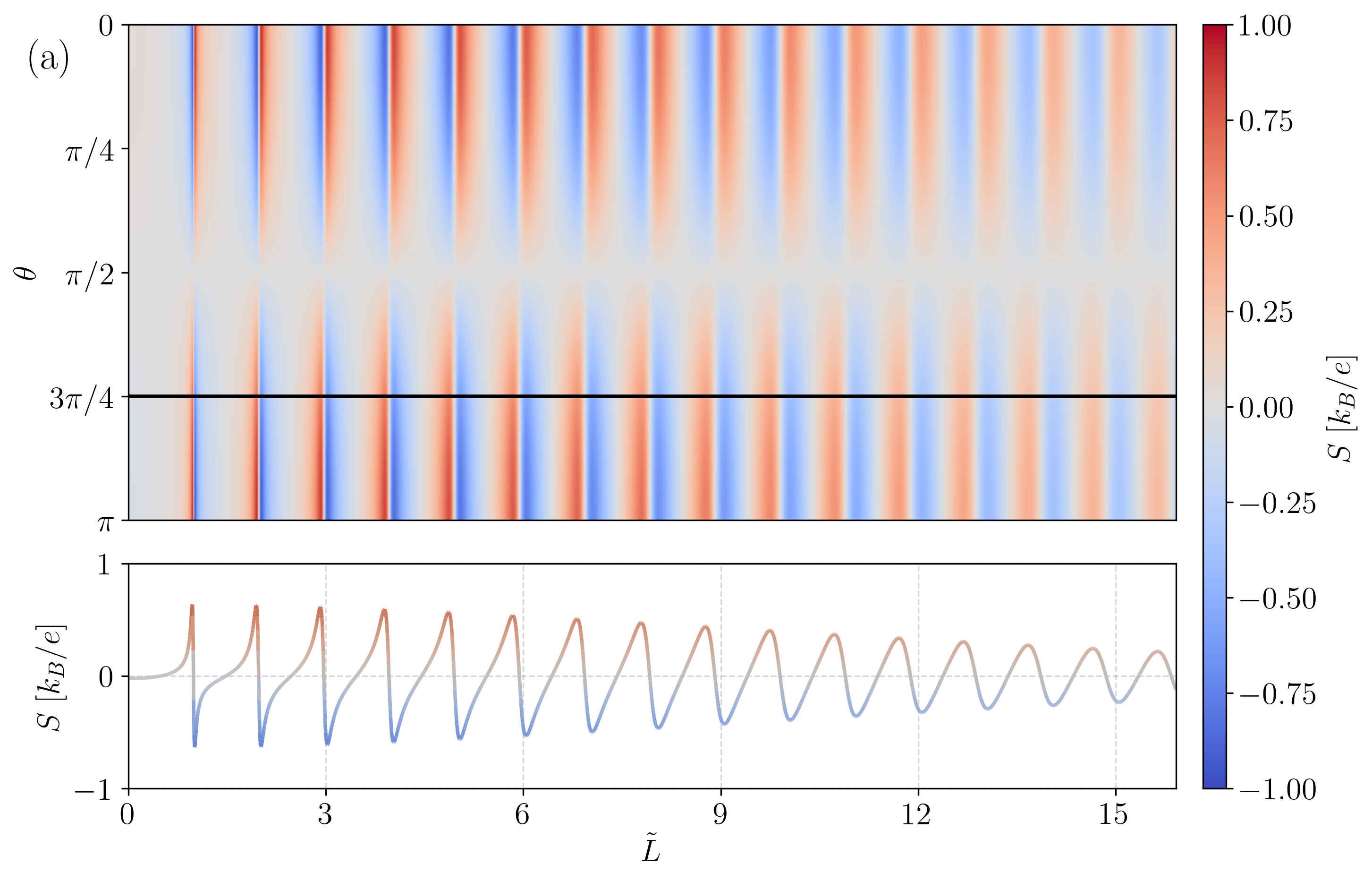}%
    \includegraphics[width=0.45\linewidth]{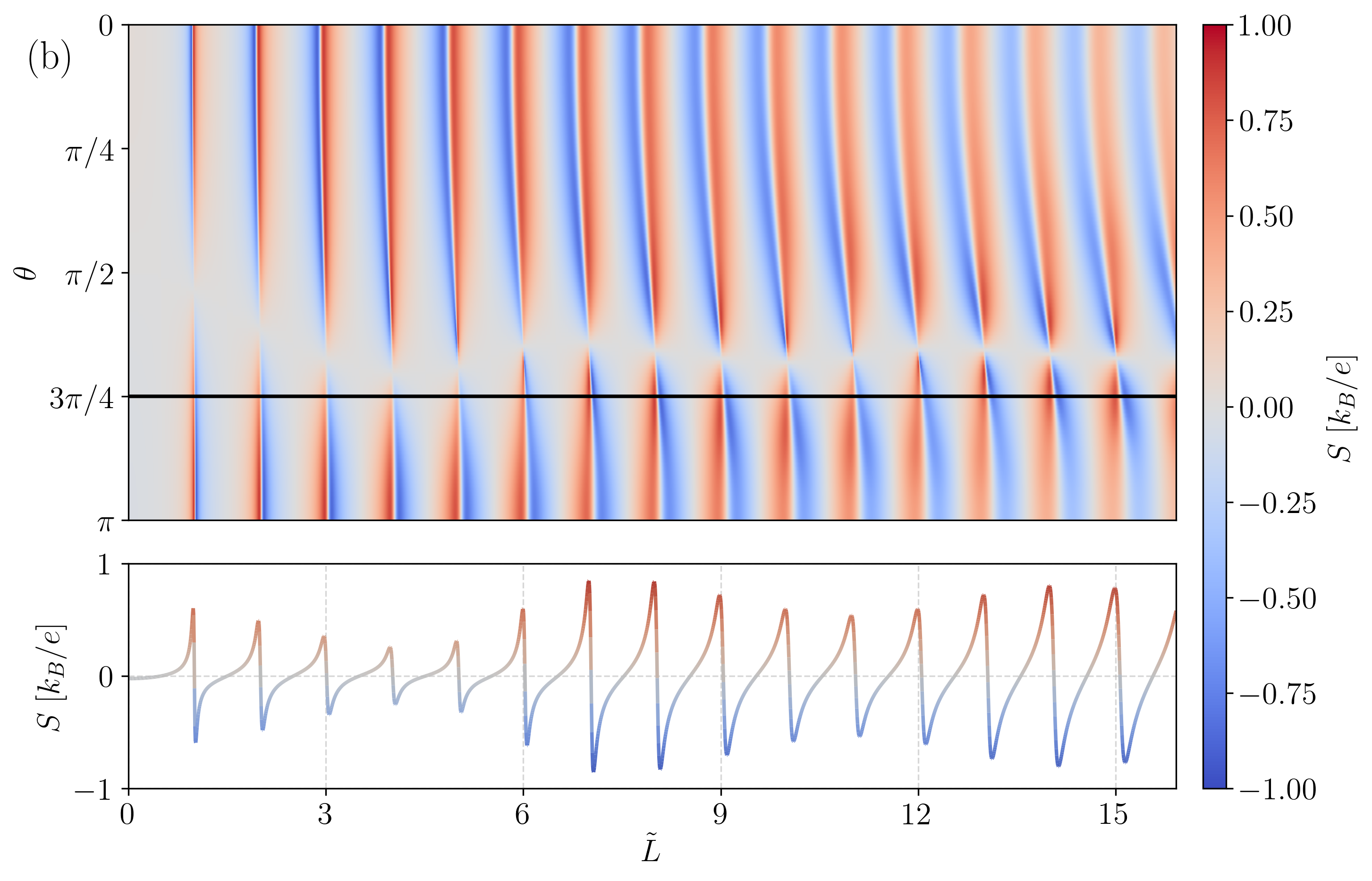}
    \caption{Thermopower density plots in the ($\Tilde{L}$-$\theta$), comparing the $\Tilde{B}=0$ (left) and $\neq 0$ (right) cases, where the plots below correspond to slices at the black lines, for $\psi=0$, $\tilde{v}_\Delta=0.1$ and $\tilde{T}=0.1$.
    It can be seen that, in the case of Zeeman splitting, there is a slight change in the $\tilde{L}$ frequency from $\theta=0$ to $\theta=\pi$, and that beatings appear in the thermopower oscillations.}
    \label{fig:elltheta}
\end{figure}

The way in which a finite Zeeman splitting breaks this symmetry is shown also in \Cref{fig:elltheta}. 
Here we show the density plot of the Seebeck coefficient as a function of $\tilde{L}$ and $\theta$. In the bottom parts of the panels, we show a cut as a function of $\tilde{L}$ for $\theta=3\pi/4$ (black line in the top panels).
The panels (a) and (b) are computed without and with the Zeeman term, respectively.
\begin{figure}
    \includegraphics[width=0.45\linewidth]{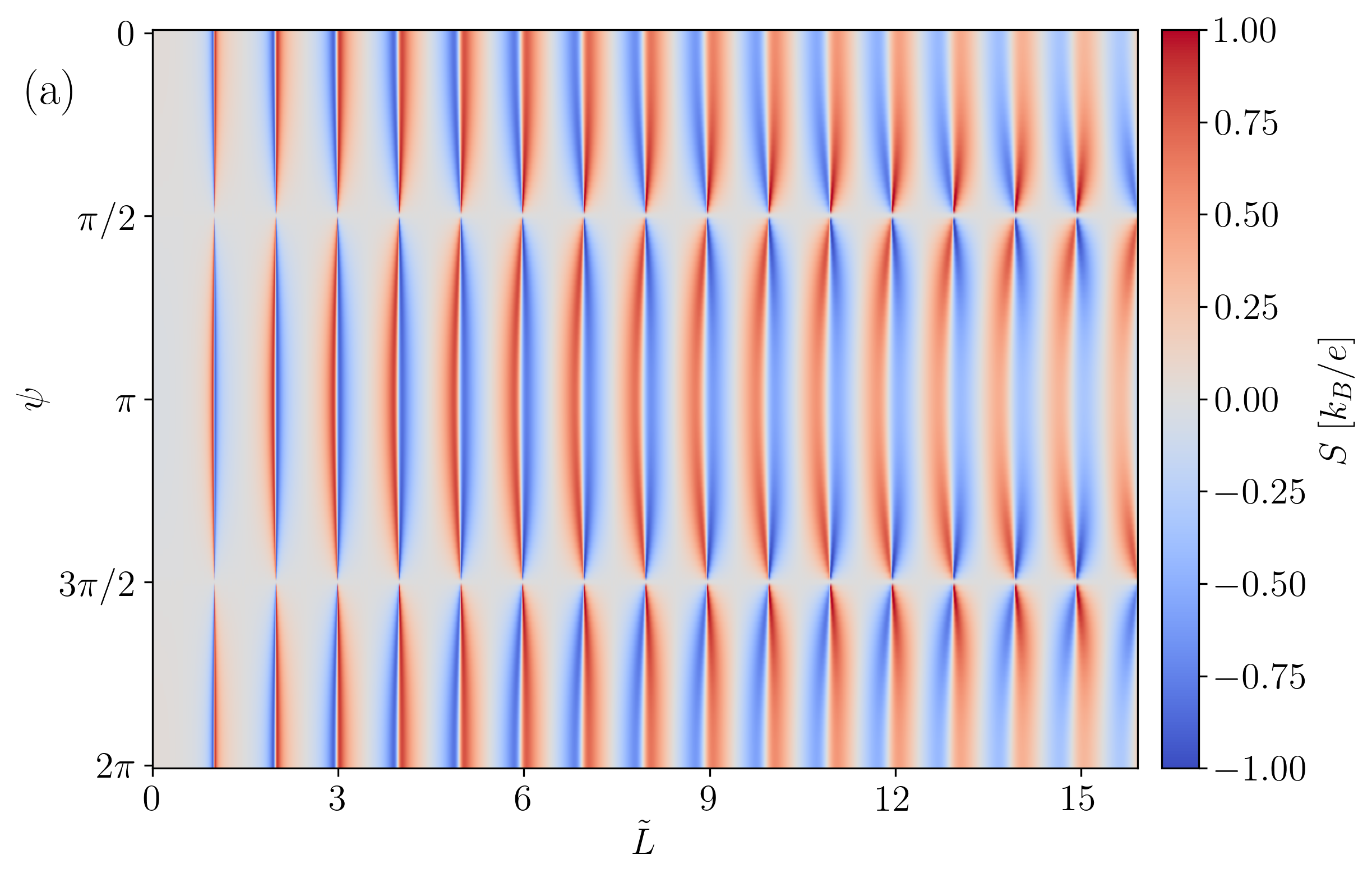}%
    \includegraphics[width=0.45\linewidth]{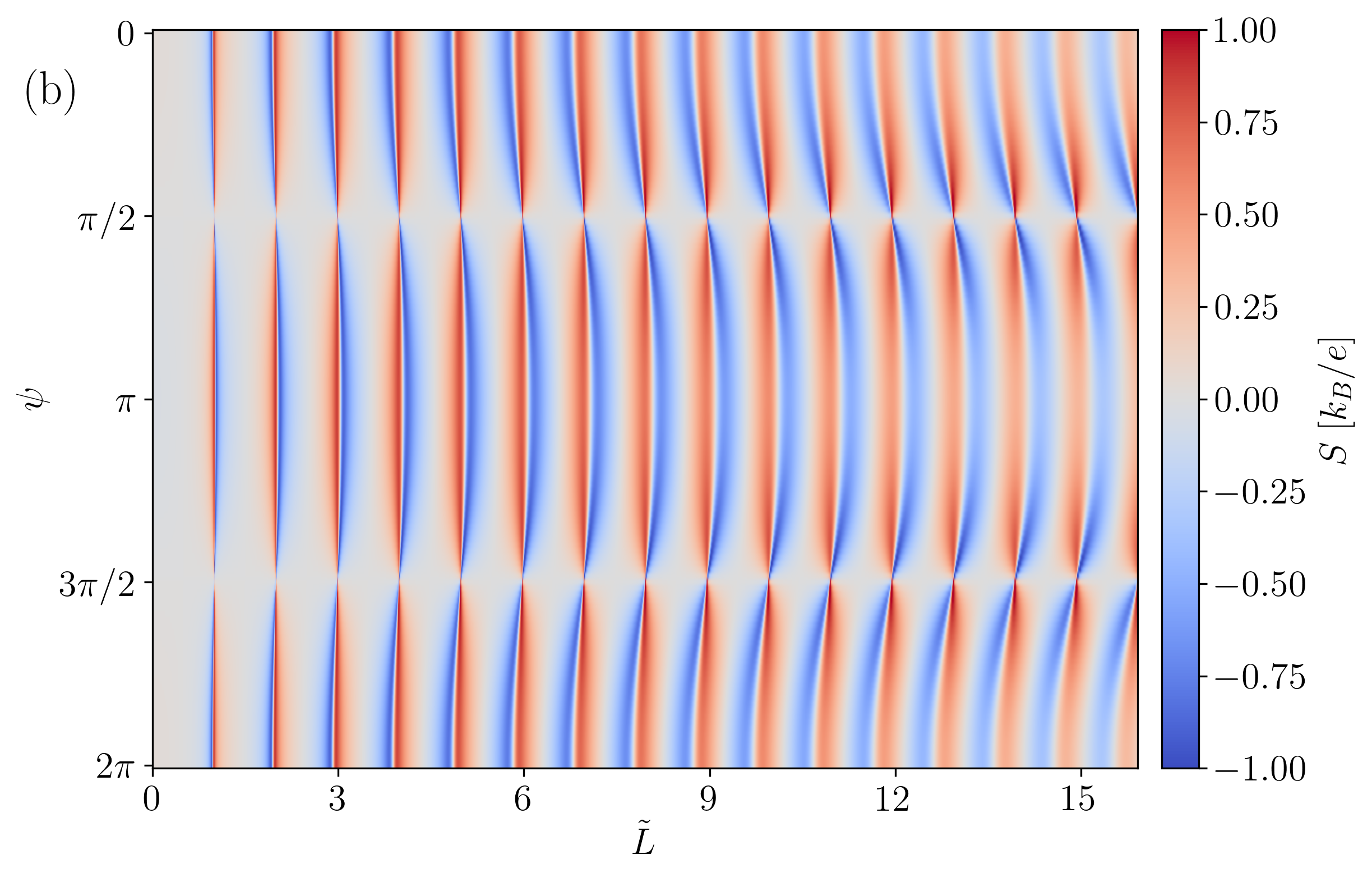}
    \caption{
    Thermopower density plots in the ($\Tilde{L}$-$\psi$) parameter space, comparing the $\Tilde{B}=0$ (left) and $\neq 0$ (right) cases, for $\theta=0$, $\tilde{v}_\Delta=0.1$ and $\tilde{T}=0.1$
    }
    \label{fig:ellpsi}
\end{figure}

The damping in $\tilde L$, for the latter case, seems to be less pronounced, with some beating behavior.
However, we also notice that panel (b) loses completely the mirror antisymmetry around $\theta=\pi/2$ present in panel (a) instead.
It is also possible to see that, with Zeeman splitting, $\theta=0$ and $\theta=\pi$ have just slightly different periodicities in $\tilde L$, and that the thermopower exhibits a beating-like behavior for different values of the angle.
\Cref{fig:ellpsi} reports the Seebeck coefficient as a function both of $\tilde{L}$ and the relative singlet-triplet phase $\psi$.
As before, panel (a) is for $\tilde{B}=0$ and panel (b) for $\tilde B\neq 0$.
Analogous to what has been observed in \Cref{fig:elltheta}, the presence of the Zeeman term slightly alters the frequency of the $S$ oscillations in $L$, also breaking the antisymmetry which would otherwise hold under the transformation $\psi\rightarrow\psi+\pi$

\subsection{Non-unitary triplet}

\begin{figure}
    \includegraphics[width=0.45\linewidth]{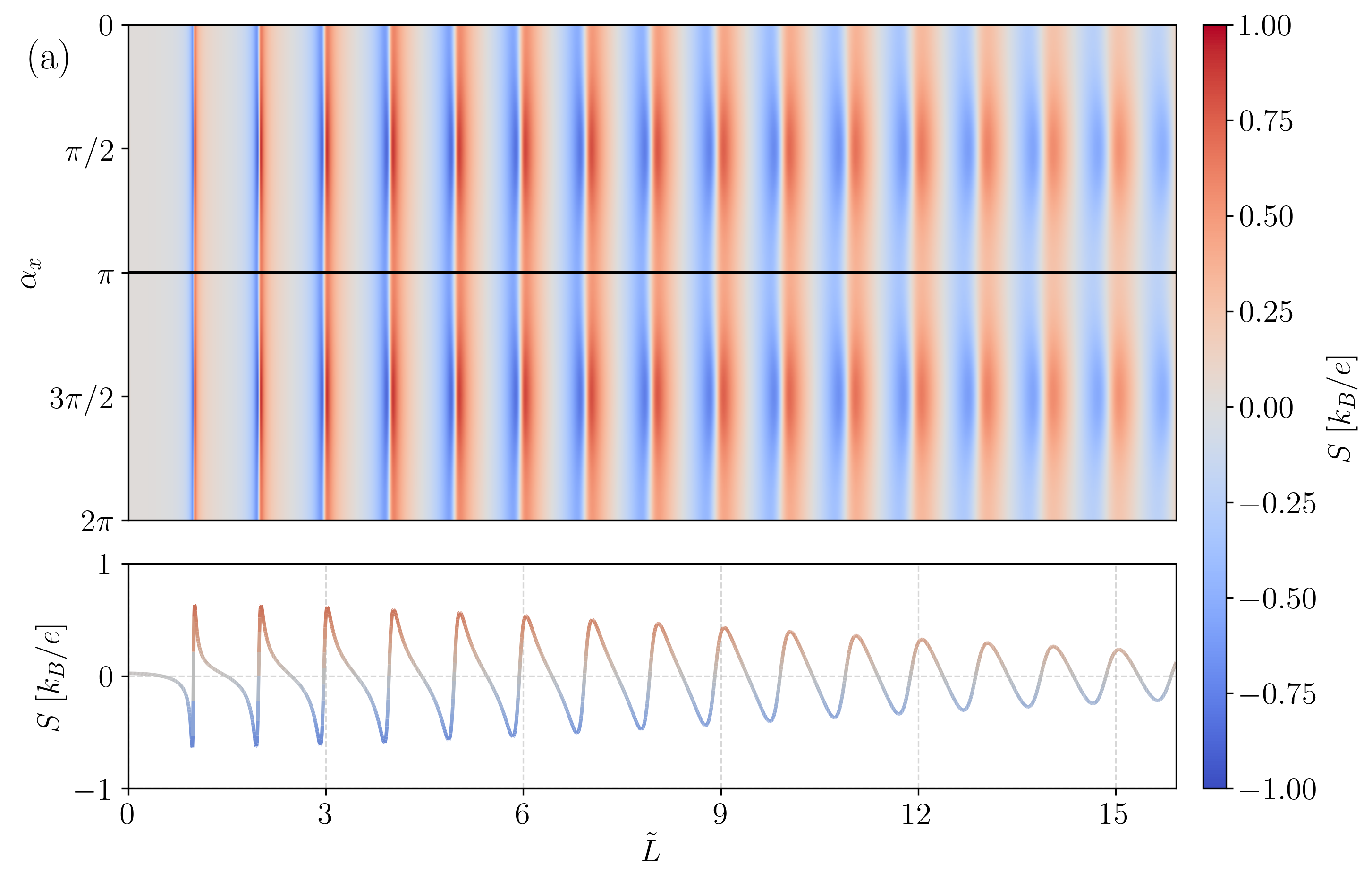}%
    \includegraphics[width=0.45\linewidth]{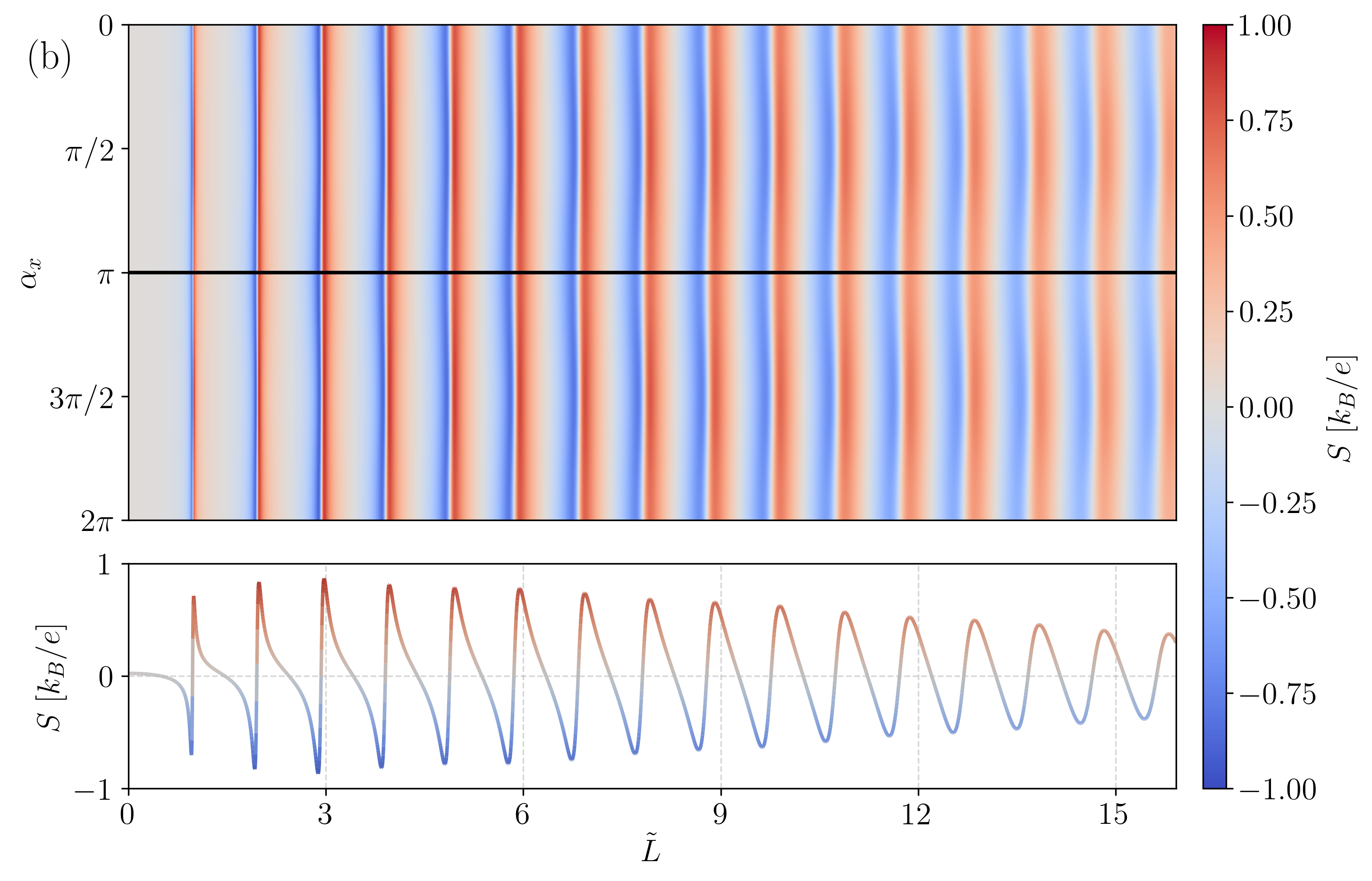}
    \caption{Thermopower density plots in the ($\Tilde{L}$-$\alpha_x$) parameter space, comparing the $\Tilde{B}=0$ (left) and $\Tilde{B}\neq 0$ (right) cases, for $\theta=0$, $\psi=0$, $\tilde{v}_\Delta=0.1$ and $\tilde{T}=0.1$.
    }
    \label{fig:ellalpha}
\end{figure}
Finally, we consider the thermopower in the more general case of a non-unitary triplet, i.e. with a non-trivial phase difference between different spatial triplet components, aiming at assessing whether this changes the TE phenomenology significantly.
In principle, the singlet ($\psi$) and the three triplet SC phases ($\alpha_x,\alpha_y,\alpha_z$)  could be all different, and, eventually, one of them could be removed via a gauge transformation.
The most general SC coupling Hamiltonian
takes the form
\begin{equation}
    \mathbf{H}_{\rm{SC}} = 
    \Delta \cos(\psi)\tau_x\sigma_0
    - \Delta \sin(\psi)\tau_y\sigma_0
    + \tilde{v}_\Delta \sum_{i\in(x,y,z)}
        \hat{v}_{\Delta,i} \left[
            \cos(\alpha_i)\tau_x\sigma_i
            - \sin(\alpha_i)\tau_y\sigma_i
        \right],
\end{equation}
where $\tilde{v}_\Delta$ still represents the normalized amplitude of the triplet vector and $\hat{v}_{\Delta,i}$ the $i$-th component of the triplet unit vector.
The quantities $\alpha_i$ are the 
complex phases of the three components.
We now consider a case entirely different from those analyzed in the main text by introducing one of the triplet SC phases.
We set $\psi=0$, since we have already analyzed the effect of changing the singlet phase, we gauge away the $\alpha_z$ triplet phase, and we set $\hat{v}_\Delta=(1,0,1)/\sqrt{2}$, so that
\begin{equation}
    \mathbf{H}_{\rm{SC}} = 
    \Delta \tau_x\sigma_0
    + \frac{\tilde{v}_\Delta}{\sqrt{2}} \left[
        \cos(\alpha_x)\tau_x\sigma_x
        - \sin(\alpha_x)\tau_y\sigma_x
        + \tau_x\sigma_z
    \right],
\end{equation}
which depends 
only on the phase $\alpha_x$.
The thermopower $S$ is numerically computed as a function of $\alpha_x$ and $\tilde L$, and plotted in \Cref{fig:ellalpha}.
This shows that in the case of non-unitary triplet pairing, the Seebeck coefficient presents the same qualitative and quantitative features as in the unitary triplet pairing case discussed in the main text. In particular, $S$ retains a resonant-like oscillating behavior in $\tilde{L}$, which is only slightly modified by $\alpha_x$.

\bibliography{references}

\end{document}